\documentclass[twocolumn,aps]{revtex4}
\usepackage{graphicx}
\usepackage{amssymb}
\usepackage{amsmath}
\usepackage{bm}
\usepackage{color}

\def\e{\varepsilon}
\def\p{\prime}
\def\s{\sigma}
\def\t{\theta}

\newcommand{\II}{\rm I\nobreak\hspace{-.1em}I}
\newcommand{\III}{\rm I\nobreak\hspace{-.1em}I\nobreak\hspace{-.1em}I}

\begin{document}

\title{Anisotropic magnetotransport in Dirac-Weyl magnetic junctions}
\author{Yuya Ominato, Koji Kobayashi, and Kentaro Nomura}
\affiliation{Institute for Materials Research, Tohoku University, Sendai 980-8577, Japan}
\date{\today}

\begin{abstract}
We theoretically study the anisotropic magnetotransport in Dirac-Weyl magnetic junctions where a doped ferromagnetic Weyl semimetal is sandwiched between doped Dirac semimetals. We calculate the conductance using the Landauer formula and find that the system exhibits extraordinarily large anisotropic magnetoresistance (AMR). The AMR depends on the ratio of the Fermi energy and the strength of the exchange interaction. The origin of the AMR is the shift of the Fermi surface in the Weyl semimetal and the mechanism is completely different from the conventional AMR originating from the spin dependent scattering and the spin-orbit interaction.
\end{abstract}
\maketitle

\section{Introduction}
\label{intro}
Magnetoresistance effects in ferromagnetic materials have been investigated during the past decades for applications to spintronics devices.
Several kinds of magnetoresistance effects have been found.
Anisotropic magnetoresistance (AMR) is a phenomenon where the resistance depends on the relative angle between the magnetization and the electric current
\cite{mcguire1975anisotropic}.
Typically the angle dependent resistivity/conductivity is of the order of a few percent.
Giant magnetoresistance (GMR) is another phenomenon
observed in thin-film structures composed of alternating ferromagnetic
and non-magnetic conductive layers
\cite{baibich1988giant,binasch1989enhanced}.
In the GMR, the magnetoresistance exhibits several dozen 
percent, which 
significantly exceeds the AMR.
Searching for 
stronger magnetoresistance effects
such as the tunneling magnetoresistance
\cite{vzutic2004spintronics}
is a central issue in the field of spintronics.

Recently, magnetotransport in topological materials,
such as topological insulators \cite{hasan2010colloquium,qi2011topological} and Dirac/Weyl semimetals
\cite{young2012dirac,wang2012dirac,murakami2007phase,wan2011topological,burkov2011weyl},
have drawn much interest for achieving novel electromagnetic coupling via the strong spin-orbit interaction
\cite{yokoyama2010anomalous,matsuzaki2015disorder,PhysRevLett.117.166806}.
A Dirac semimetal manifests the pseudo-relativistic linear dispersions doubly degenerate with time-reversal and spatial-inversion symmetries, and it is non-magnetic
\cite{liu2014discovery,neupane2014observation,borisenko2014experimental}.
A Weyl semimetal possesses the gapless linear dispersions with broken time-reversal \cite{liu2015discovery,wang2016large,borisenko2016time} or spatial-inversion symmetries
\cite{xu2015discovery,lv2015experimental,xu2015discovery2,belopolski2015observation,yang2015weyl,xu2015experimental,xu2016observation}.
In a magnetic Weyl semimetal, time reversal symmetry is broken because of a magnetic ordering.
One of the signatures of Dirac/Weyl semimetals is the longitudinal negative magnetoresistance
\cite{nielsen1983adler,aji2012adler,son2013chiral,burkov2014chiral,gorbar2014chiral,lu2015high,burkov2015negative} that has been observed experimentally in non-magnetic systems
\cite{xiong2015evidence,li2015giant,zhang2015detection,huang2015observation,zhang2015observation,yang2015chiral,li2016negative,arnold2016negative,zhang2016signatures,dos2016search}.
The negative magnetoresistance can be observed in both time-reversal and spatial-inversion symmetry broken Weyl semimetals. Here, we study magnetoresistance effects that is peculiar to the time-reversal symmetry broken Weyl semimetals, i.e., magnetic Weyl semimetals.


In this paper, we study the anisotropic magnetotransport in 
Dirac-Weyl magnetic junctions.
We consider a system that consists of a doped ferromagnetic Weyl semimetal sandwiched by doped 
Dirac semimetals, and calculate the transmission probability
through the Dirac-Weyl magnetic junctions. 
In our system, the Dirac/Weyl semimetals are 
characterized by the absence/presence of the spontaneous magnetization, which splits the band touching points in momentum space.
Using the Landauer formula, we compute the conductance as a function of the relative angle between the magnetization and the electric current.
We find that the AMR can be extraordinarily large compared with that of conventional ferromagnetic metals and the AMR has the same periodicity of the conventional AMR as a function of the relative angle.
In the Dirac-Weyl magnetic junctions,
the mechanism of the AMR is
entirely 
different from
the conventional one,
which is generally considered to arise from spin dependent scattering and spin-orbit coupling.
The shift of the Fermi surface
caused by the magnetization in the 
Weyl semimetal 
is the origin of the AMR.

The paper is organized as follows.
In Sec.~\ref{sec_model}, we introduce a continuous model
which describes a Dirac semimetal and a ferromagnetic Weyl semimetal with a pair of Weyl nodes.
In Sec.~\ref{sec_trans},
we calculate the transmission probability of the Dirac-Weyl magnetic junctions.
In Sec.~\ref{sec_cond}, we compute the conductance using the Landauer formula.
The conclusion and discussion are given in Sec.~\ref{sec_conc}.

\section{Model Hamiltonian}
\label{sec_model}

\begin{figure}
\begin{center}
\leavevmode\includegraphics[width=0.8\hsize]{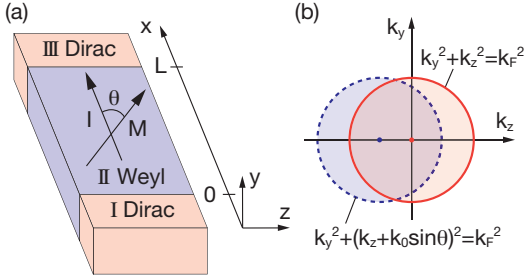}
\end{center}
\caption{(a) Dirac-Weyl 
magnetic junctions composed of Dirac and
ferromagnetic Weyl semimetals. The magnetization vector
is on the $x$-$z$ plane and the angle between the magnetization
and the $x$ axis is $\t$.
(b) Fermi surfaces of Dirac and Weyl semimetals with positive chirality
projected on $k_y$-$k_z$ plane.
The area inside the red solid 
(blue dashed) 
circle represents the Fermi surface of the Dirac 
(Weyl) 
semimetal.
}
\label{fig_junc}
\end{figure}

We consider magnetic junctions composed of Dirac
and ferromagnetic Weyl semimetals as illustrated in
Fig.~\ref{fig_junc}(a), where the doped magnetic Weyl semimetal
in region {\II} $(0\le x\le L)$
is sandwiched between the doped Dirac semimetals
in region I and {\III} $(x<0, L<x)$.
We assume that the Dirac semimetals are semi-infinite in $x$ direction
and the system is periodic in $y$ and $z$ direction.
The magnetic junction
is implemented in the magnetization,
\begin{equation}
\bm{M}(x)=\begin{cases}
                 \hspace{5mm}(0,0,0) \hspace{15mm}(x<0, L<x),\\
                 M_0(\cos\t,0,\sin\t) \hspace{5mm}(0\le x\le L),
                 \end{cases}
\end{equation}
as shown in Fig.~\ref{fig_junc}(a).

We start with a model Hamiltonian for electrons in Dirac/Weyl semimetals,
\begin{align}
H_0 = \hbar v\tau_z\bm{\s}\cdot(-i\bm{\nabla})+\tau_0J\bm{M}(x)\cdot\bm{\sigma},
\end{align}
where $v$ is the velocity,
$\bm{\s}$ and $\bm{\tau}$ are the 
triplets 
of Pauli matrices acting on the real spin and the pseudospin (chirality) degrees of freedom,
and $J$ is the exchange coupling constant.
There are two nodes with positive and negative chirality characterizing the correlation between the spin and the momentum.
In region I and {\III} (doped Dirac semimetal), there is no magnetic moment, and the energy bands are doubly degenerate.
In region {\II} (doped Weyl semimetal), the exchange interaction splits the two-fold Fermi surfaces in the direction of the magnetization in momentum space.
In the present work, we treat two Fermi surfaces with positive and negative 
chiralities 
independently,
assuming the absence of inter-node scattering.
Figure~\ref{fig_junc}(b) shows the Fermi surfaces of the Dirac and Weyl semimetals with positive chirality,
projected on $k_y$-$k_z$ plane.
In the presence of the magnetization
(in region {\II}),
the Fermi surface 
is shifted along the $k_z$ axis by $k_0\sin\t$, where
\begin{align}
   k_0={JM_0\over \hbar v}.
\end{align}
When $k_{\rm F}>k_0\sin\t$, where $k_{\rm F}$ is the Fermi wave number,
the projected Fermi surfaces of the Dirac and Weyl semimetals 
are partially overlapped, while not when $k_{\rm F}<k_0\sin\t$.

\section{Transmission probability}
\label{sec_trans}

In this section, we calculate transmission probability along $x$ axis
for eigenstates of the positive chirality.
Due to the translational invariance along $y$ and $z$ axis,
the wave numbers $k_y$ and $k_z$ are conserved and
an eigenstate is labeled by the wave numbers $k_y$ and $k_z$
in the projected Fermi surface of the Dirac 
semimetals,
$k_y^2+k_z^2<k_{\rm F}^2$.
The common factor $e^{i(k_y y+k_z z)}$ 
is 
omitted
in the following expressions.

The wave function can be written in terms of incident and reflected waves.
In region I, the two-component wave function is written as
\begin{align}
\psi_{\rm I}(x)=&\begin{pmatrix}
                                   k_x-ik_y \\
                                   k_{\rm F}-k_z
                        \end{pmatrix}e^{ik_x x} \notag \\
                        &+r\begin{pmatrix}
                                   -k_x-ik_y \\
                                   k_{\rm F}-k_z
                         \end{pmatrix}e^{-ik_x x},
\label{wfI}
\end{align}
where
\begin{align}
k_x=\sqrt{k_{\rm F}^2-k_y^2-k_z^2}\ .
\label{kx}
\end{align}
In region \II, we have
\begin{align}
\psi_{\II}(x)=&\hspace{4mm}a\begin{pmatrix}
                                   k_x^\p-ik_y \\
                                   k_{\rm F}-(k_z+k_0\sin\t)
                                 \end{pmatrix}e^{i(k_x^\p -k_0\cos\t)x} \notag \\
                        &+b\begin{pmatrix}
                                   -k_x^\p-ik_y \\
                                   k_{\rm F}-(k_z+k_0\sin\t)
                            \end{pmatrix}e^{i(-k_x^\p-k_0\cos\t)x},
\label{wfII}
\end{align}
where
\begin{align}
k_x^\p=\sqrt{k_{\rm F}^2-k_y^2-(k_z+k_0\sin\t)^2}\ .
\label{kxp}
\end{align}
In the case of
$k_{\rm F}^2-k_y^2-(k_z+k_0\sin\t)^2<0$,
$k_x^\p$ becomes pure imaginary number,
and the wave function describes an evanescent mode decaying exponentially.
In region \III, we have
\begin{align}
\psi_{\III}(x)=t\begin{pmatrix}
                                   k_x-ik_y \\
                                   k_{\rm F}-k_z
                                 \end{pmatrix}e^{ik_x x}.
\label{wfIII}
\end{align}

The continuity of the wave functions at the junctions
gives boundary conditions,
$\psi_{\rm I}(0)=\psi_{\II}(0)$ and
$\psi_{\II}(L)=\psi_{\III}(L)$,
and 
determines 
the coefficients $r$, $a$, $b$, and $t$.
The transmission probability along $x$ axis is obtained from $T(k_y,k_z,\t)=|t|^2$
and has the form
\begin{align}
&T(k_y,k_z,\t)= \notag \\
&\frac{4k_x^2 k_x^{\p\,2}}{4k_x^2 k_x^{\p\,2} \cos^2(k_x^\p L)
 +\left(k_x^2\!+\!k_x^{\p\,2}\!+\!k_0^2\sin^2\!\t\right)^{\!2}\!\sin^2(k_x^\p L)}.
\label{eq_trans}
\end{align}
Note that $k_x$ and $k_x^\p$ are functions of
$k_y$ and $k_z$ as given in 
Eqs.~(\ref{kx}) and (\ref{kxp}).
From the above expression, we see that the transmission probability
depends on
the $z$ component of the magnetization,
giving the shift of the 
projected Fermi surface, 
while 
the $x$ component 
does not contribute. 
Therefore, the transmission probability is unity
at $\t=m\pi$ with an integer $m$.

As shown in Fig.~\ref{fig_junc}(b),
the projected Fermi surface of the Dirac semimetals ($k_y^2+k_z^2<k_{\rm F}^2$) can be divided into two regions:
overlapping region [$k_y^2\!+\!(k_z\!+\!k_0\sin\!\t)^2\!<\!k_{\rm F}^2$], 
where that of the Weyl semimetal is overlapping, 
and non-overlapping region [$k_y^2+(k_z+k_0\sin\t)^2\!>\!k_{\rm F}^2$], where the traveling mode corresponding to the incident wave is absent in the Weyl semimetal.
The expression Eq.~(\ref{eq_trans})
is applicable 
also 
to 
a 
pure imaginary $k_x^\p$,
i.e., the wave function in region {\II} is an evanescent mode.
The transmission probability behaves in a significantly different manner
in the overlapping (with real $k_x'$) and non-overlapping (with pure imaginary $k_x'$) regions.

\begin{figure}[t]
\begin{center}
\leavevmode\includegraphics[width=0.95\hsize]{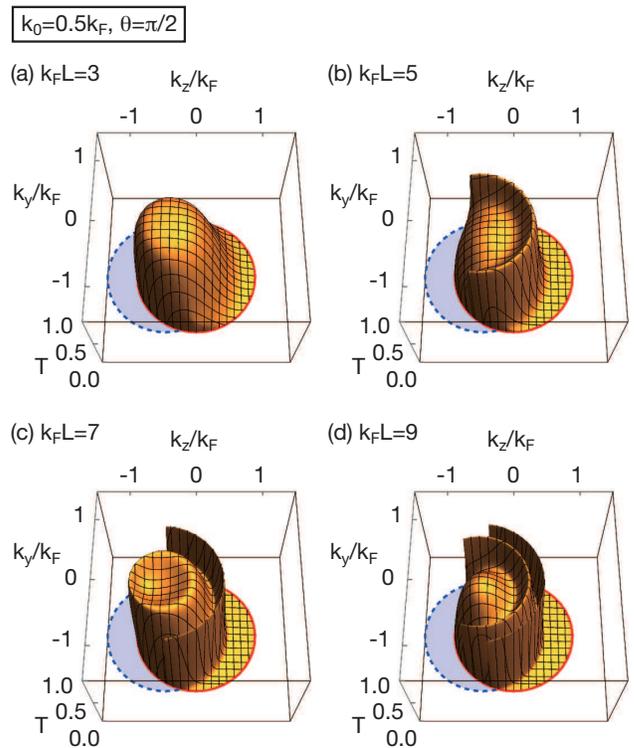}
\end{center}
\caption{Transmission probability as a function of $k_y$ and $k_z$
in the Fermi surface of the Dirac semimetals projected on the $k_y$-$k_z$ plane.
Transmission probability is plotted at (a) $k_{\rm F} L=3$,
(b) $k_{\rm F} L=5$, (c) $k_{\rm F} L=7$, (d) $k_{\rm F} L=9$.
We set $k_0=0.5k_{\rm F}$ and $\t=\pi/2$.
}
\label{fig_trans_length}
\end{figure}

In the overlapping region, the incident wave is transmitted
via the traveling mode in region \II.
From Eq.~(\ref{eq_trans}), we can show that the incident wave is transmitted
with the transmission probability $T(k_y,k_z,\t)=1$ for values of $k_x^\p L$
satisfying the relation $k_x^\p L=n\pi$
\cite{bai2016wavevector,bai2016chiral,you2016electronic,yesilyurt2016klein,siu2016influence}, with $n$ a positive integer,
corresponding to a condition that a standing wave
can exist in the region \II.
The relation is written as
\begin{align}
k_y^2+(k_z+k_0\sin\t)^2=k_{\rm F}^2\left[1-\left(\frac{n\pi}{k_{\rm F}L}\right)^2\right].
\label{eq_circle}
\end{align}
Figure \ref{fig_trans_length} shows the transmission probability as a function of $k_y$ and $k_z$
at several $L$'s. We set $k_0=0.5k_{\rm F}$ and $\t=\pi/2$ as a typical example.
For $k_{\rm F}L<\pi$,
there is no solution 
of Eq.~(\ref{eq_circle}), 
so that there is no peak structure
in Fig.~\ref{fig_trans_length}(a) and the transmission probability satisfies $T(k_y,k_z,\t)<1$.
In Figs.~\ref{fig_trans_length}(b), (c), and (d), 
there are peak structures 
where the transmission probability is unity
on the circles represented by Eq.~(\ref{eq_circle}).
The number of peaks increases with the increase of $k_{\rm F}L$.

In the non-overlapping region,
an evanescent mode in region \II,
$k_x^\p=i\kappa$ with $\kappa=\sqrt{-k_{\rm F}^2+k_y^2+(k_z+k_0\sin\t)^2}$,
transmits an incident wave and the transmission probability is written as
\begin{align}
 &T(k_y,k_z,\t)= \notag \\
 &\frac{4k_x^2{\kappa}^2}{
    4k_x^2{\kappa}^2\cosh^2(\kappa L)
   +\left(k_x^2\!-\!\kappa^2\!+\!k_0^2\sin^2\!\t\right)^2\!\sinh^2(\kappa L)
  }.
\label{eq_evanes}
\end{align}
From the above expression, we immediately see that
the transmission probability monotonically decreases with the increase of $L$ in an exponential manner.
Figure \ref{fig_trans_length} shows that the transmission probability becomes exponentially small.

\begin{figure}[t]
\begin{center}
\leavevmode\includegraphics[width=0.95\hsize]{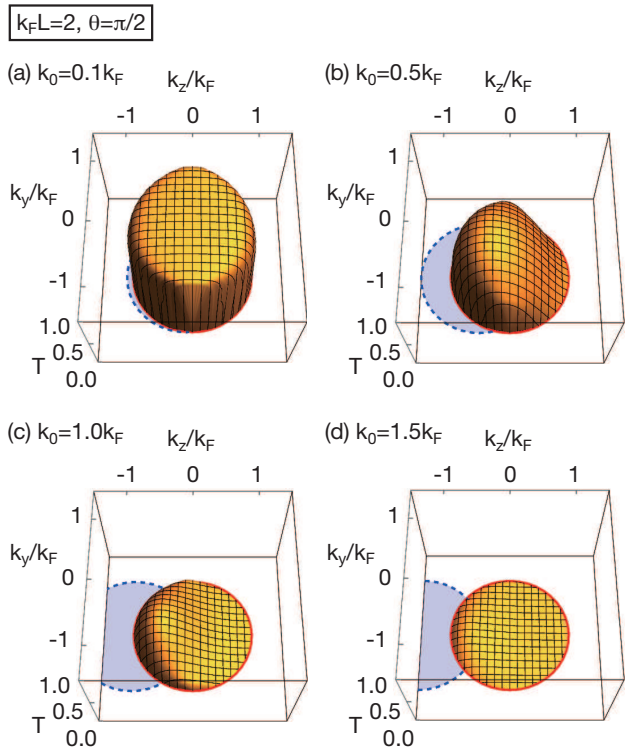}
\end{center}
\caption{Transmission probability as a function of $k_y$ and $k_z$
in the Fermi surface of the Dirac semimetals projected on the $k_y$-$k_z$ plane.
Transmission probability is plotted at (a) $k_0=0.1k_{\rm F}$,
(b) $k_0=0.5k_{\rm F}$, (c) $k_0=1.0k_{\rm F}$, (d) $k_0=1.5k_{\rm F}$.
We set $k_{\rm F} L=2$ and $\t=\pi/2$.
}
\label{fig_trans_k0}
\end{figure}

Figure \ref{fig_trans_k0} shows the transmission probability at several $k_0$'s,
where we set $k_{\rm F}L=2$ and $\t=\pi/2$.
For a large $k_0$, the transmission probability is significantly suppressed even in the overlapping region; the spin-momentum locking causes the discrepancy of the spinors between the incident and traveling modes.

\section{Conductance}
\label{sec_cond}

To characterize the total transmission for 
the 
given magnetization direction and magnitude,
we calculate the Landauer conductance
\begin{align}
G(\t)=&2\frac{e^2}{h}\sum_{k_y k_z}T(k_y, k_z,\t) \notag \\
       =&2\frac{e^2}{h}\frac{S}{(2\pi)^2}\int_D T(k_y, k_z,\t) dk_y dk_z,
\label{eq_cond}
\end{align}
where $S$ is the area of the junctions, and
the region $D$ is the projected Fermi surface of the Dirac semimetals, $k_y^2+k_z^2<k_{\rm F}^2$.
The factor two comes from the
two nodes
with positive and negative chiralities giving the same contribution.
The conductance $G(\t)$ is naturally proportional to $S$,
and the conductance per unit area, $G(\t)/S$,
can be used as a quantity that measures the transparency of
the magnetic junctions for the electronic transport.

\begin{figure}
\begin{center}
\leavevmode\includegraphics[width=1.0\hsize]{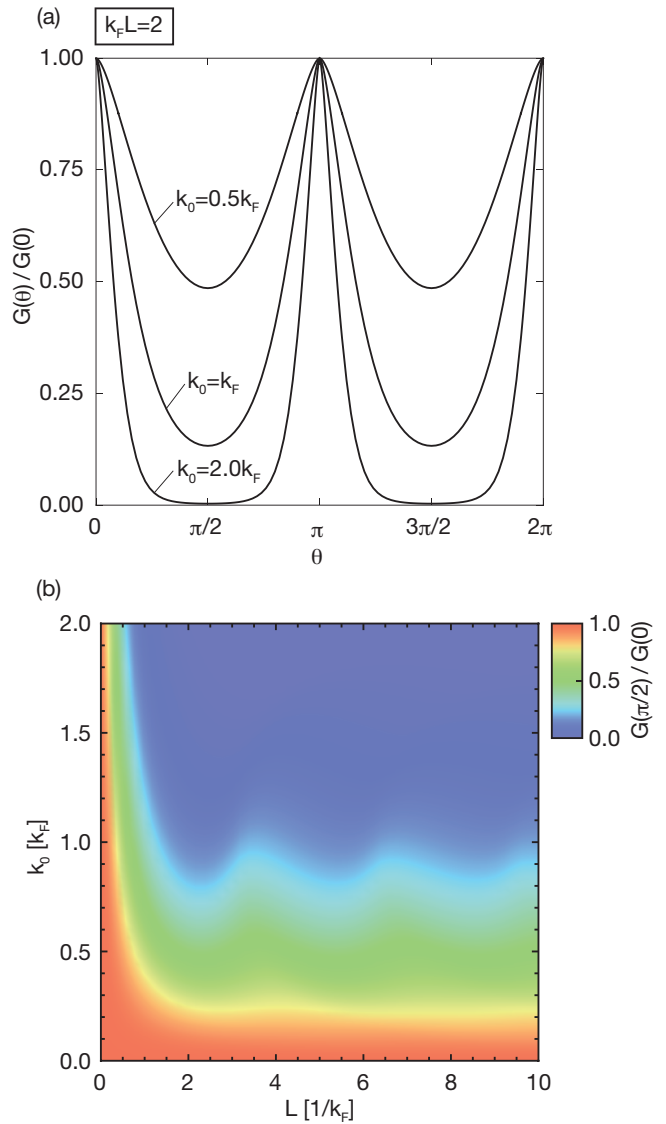}
\end{center}
\caption{(a) The conductance $G(\t)$ as a function of $\t$.
We set $k_{\rm F}L=2$ and plot the conductance
for $k_0=0.5k_{\rm F}, k_{\rm F}, 2.0k_{\rm F}$.
(b) The color plot of the ratio of the conductance 
$G(\pi/2)/G(0)$ 
as a function of $L$ and $k_0$.}
\label{fig_cond}
\end{figure}

In Fig.~\ref{fig_cond}(a), we plot the conductance, Eq.~(\ref{eq_cond}), as a function of $\t$.
Here, we set $k_{\rm F}L=2$ and $k_0=0.5k_{\rm F}$, $k_{\rm F}$, $2k_{\rm F}$.
The conductance has a property, $G(\t+\pi)=G(\t)$,
i.e., $G(\t)$ has a periodicity of $\pi$,
which can be derived from Eq.~(\ref{eq_trans}), the expression for $T(k_y,k_z,\t)$,
and appropriately changing the integral variable in Eq.~(\ref{eq_cond}).
At the angles $\t=m\pi$ with an integer $m$,
the conductance is independent of $k_0$ and $\t$, because
the transmission probability is unity for the arbitrary incident wave
as we mentioned in the previous section.
The conductance decreases when the angle $\t$ deviates from $\t=m\pi$
and reaches a minimum value at $\t=(m+1/2)\pi$.
This is because the conductance is governed by the area of the overlapping region, and the overlapping area depends only on the shift of the projected Fermi surface $|k_0\sin\t|$. Therefore, the conductance decreases with the increase of $k_0$, as we can see it in Fig.~\ref{fig_cond}(a). At $k_0=2k_{\rm F}$, there is no overlapping region at $\t=(m+1/2)\pi$, so that the conductance becomes exponentially small.

In Fig.~\ref{fig_cond}(b), the conductance $G(\pi/2)$, which is the minimum value of $G(\t)$, is plotted as a function of $L$ and $k_0$. When $k_0$ is fixed, the conductance decreases with the increase of $L$ because the transmission probability in the non-overlapping region decreases exponentially, and approaches to a minimum value, which is approximately proportional to the area of the overlapping region. Around $k_0\approx0.7k_{\rm F}$, we see oscillatory behavior coming from the peak structures of the transmission probability. The conductance at fixed $L$ decreases with the increase of $k_0$ because of the decrease of the overlapping area.


\section{Conclusion and Discussion}
\label{sec_conc}

We studied the anisotropic magnetotransport in the Dirac-Weyl magnetic junctions
and found that the present system exhibits the extraordinarily large AMR.
For the magnetization parallel to the electric current ($\t=0$),
the conductance is not influenced by the magnetization,
while for the magnetization perpendicular to the electric current ($\t=\pi/2$),
the conductance becomes exponentially small for a sufficiently strong exchange interaction and magnetization, i.e., $k_0\gg k_{\rm F}$.

\begin{figure}[t]
\begin{center}
\leavevmode\includegraphics[width=0.95\hsize]{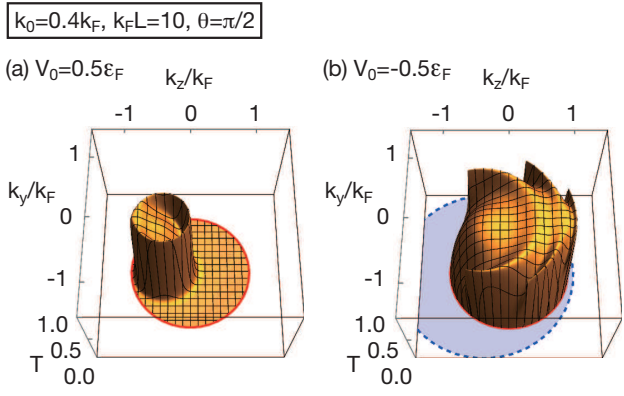}
\end{center}
\caption{
Transmission probability as a function of $k_y$ and $k_z$
in the Fermi surface of the Dirac semimetals projected on the $k_y$-$k_z$ plane.
Transmission probability is plotted at (a) $V_0=0.5\e_{\rm F}$ and
(b) $V_0=-0.5\e_{\rm F}$, where $\e_{\rm F}=\hbar v k_{\rm F}$.
We set $k_0=0.4k_{\rm F}$, $k_{\rm F} L=10$, and $\t=\pi/2$.
}
\label{fig_trans_v0}
\end{figure}

Here we discuss the case when the sizes of the Fermi surface in the Dirac and Weyl semimetals are different, and show that the difference does't change the qualitative results. In this paper, we have assumed that the sizes of the Fermi surface are the same, although they can be different in general. In this case, the Hamiltonian is given as
\begin{align}
H=H_0+V(x),
\end{align}
where $V(x)$ is a potential, shifting the Fermi energy of the Weyl semimetal,
\begin{equation}
V(x)=\begin{cases}
                 0 \hspace{4.8mm}(x<0, L<x),\\
                 V_0 \hspace{3mm}(0\le x\le L).
                 \end{cases}
\end{equation}
In a similar manner to Eq.~(\ref{eq_trans}),
we derive the transmission probability,
\begin{align}
&T(k_y,k_z,\t)= \notag \\
&\hspace{-4mm}\frac{4k_x^2 k_x^{\p\,2}}{4k_x^2 k_x^{\p\,2} \cos^2(k_x^\p L)
 +\left[k_x^2\!+\!k_x^{\p\,2}\!+\!k_0^2\sin^2\!\t-V_0^2/(\hbar v)^2\right]^{\!2}\!\sin^2(k_x^\p L)},
\label{eq_trans2}
\end{align}
where
\begin{align}
k_x^\p&=\sqrt{\left[k_{\rm F}-V_0/(\hbar v)\right]^2-k_y^2-(k_z+k_0\sin\t)^2}\ .
\label{kxp2}
\end{align}
In Fig.\ \ref{fig_trans_v0}, the transmission probability is plotted
at (a) $V_0=0.5\e_{\rm F}$ and (b) $V_0=-0.5\e_{\rm F}$,
where $\e_{\rm F}=\hbar v k_{\rm F}$.
We again observe the peak structures on $k_x^\p L=n\pi$
and the suppression of the transmission probability
originating from the spin-momentum locking.

The origin of the AMR in the present system is the shift of the Fermi surface in the momentum space, which is completely different from that of the conventional AMR in ferromagnetic metal, but we can get an expression for the conductance that resembles the conventional AMR.
On condition that $|k_0\sin\t|$ is much smaller than
the Fermi wave number $k_{\rm F}$,
the transmission probability Eq.\ (\ref{eq_trans}) is approximated by
\begin{align}
T(k_y,k_z,\t)\simeq 1-T_1&(k_y,k_z)\left(\frac{k_0\sin\t}{k_{\rm F}}\right) \notag \\
                                         &-T_2(k_y,k_z)\left(\frac{k_0\sin\t}{k_{\rm F}}\right)^2,
\label{eq_app_trans}
\end{align}
where we neglect higher order terms than $(k_0\sin\t/k_{\rm F})^2$. Substituting Eq.~(\ref{eq_app_trans}) for Eq.~(\ref{eq_cond}), we get an approximate expression. The conductance $G(\t)$ is an even function of $\t$, i.e., $G(\t)=G(-\t)$, because the system with the relative angle $-\t$ can be transformed into the relative angle $\t$ by rotating the system around $x$ axis. Therefore, the conductance is written as
\begin{align}
G(\t)\simeq G(0)-\Delta G\sin^2\t,
\end{align}
where
\begin{align}
&G(0)=2\frac{e^2}{h}\frac{S}{(2\pi)^2}\pi k_{\rm F}^2, \notag \\
&\Delta G=2\frac{e^2}{h}\frac{S}{(2\pi)^2}\int_DT_2(k_y,k_z)\left(\frac{k_0}{k_{\rm F}}\right)^2dk_ydk_z.
\end{align}
The above expression for $G(\t)$ is similar to the AMR in the conventional ferromagnetic metals \cite{mcguire1975anisotropic}.

Finally, we mention how to observe the extraordinarily large AMR experimentally.
There is a great deal of theoretical and experimental work on searching for magnetic Weyl semimetals.
One of the candidate materials is Cr-doped ${\rm Bi}_2({\rm Se}_x{\rm Te}_{1-x})_3$
\cite{chen2010massive,zhang2013topology,kurebayashi2014weyl}.
The bulk band gap of ${\rm Bi}_2{\rm Se}_3$ can be tuned by substituting tellurium for selenium, and
the gap almost closes at the $\Gamma$ point with the selenium concentration $x\simeq0.6$
\cite{sato2011unexpected,jin2011topological}.
In the presence of magnetic dopants Cr, we expect the ferromagnetic ordering
below a critical temperature and emergence of the Weyl semimetal phase.
Therefore, the AMR discussed in the present work can be observed
in multilayer structure of non-magnetic and magnetic ${\rm Bi}_2({\rm Se}_x{\rm Te}_{1-x})_3$.
The condition to observe the large AMR is written as $k_{\rm F}\ll k_0$ and $L\gg 1/k_{\rm F}$.
Using the parameters for the above material
($JM_0=2.0 x_i S[{\rm eV}], \hbar v=2.2[{\rm eV \AA}]$)
\cite{yu2010quantized,chen2010massive,zhang2013topology,zhang2009topological,liu2010model},
where $x_i$ is the ratio of the magnetic dopants and $S=3/2$,
we can quantitatively estimate the shift of the Fermi surface $k_0$.
At $x_i=0.1$, the condition is satisfied when
$k_{\rm F}\ll 0.14[{\rm \AA}^{-1}] (\e_{\rm F}\ll 0.31[{\rm eV}])$.
In this situation, we can set $k_{\rm F}=0.01[{\rm \AA}^{-1}]$ as a typical value, so that
the condition for the system size is written as $L\gg 100[{\rm \AA}]$.

\section*{ACKNOWLEDGEMENTS}

This work was supported by JSPS KAKENHI Grant Numbers JP26400308, JP15H05854, and JP16J01981.

\bibliography{dirac_weyl_magnetic_junction}

\end{document}